\documentclass[superscriptaddress,aps,prb,citeautoscript,  preprint, amsmath, amssymb]{revtex4-2}

\usepackage{bm}
\usepackage{amsmath}
\usepackage{graphicx}
\usepackage{amssymb}
\usepackage{dcolumn}
\usepackage{bm}
\usepackage{lipsum}
\usepackage{gensymb}
\usepackage{siunitx}
\usepackage{tabularx}
\usepackage{multirow}
\usepackage{setspace}

\usepackage{booktabs}
\usepackage{natbib}
\usepackage[dvipsnames]{xcolor}

\usepackage[english]{babel}

\DeclareUnicodeCharacter{2212}{-}

\newcommand{\rpm}{\sbox0{$1$}\sbox2{$\scriptstyle\pm$}\raise\dimexpr(\ht0-\ht2)/2\relax\box2 }

\begin{document}

\preprint{APS/123-QED}

\title{Low Thermal Resistance of Diamond-AlGaN Interfaces Achieved Using Carbide Interlayers}

\author{Henry T. Aller}
\affiliation{Department of Mechanical Engineering, University of Maryland, College Park, Maryland 20742, USA}
\author{Thomas W. Pfeifer}
\affiliation{Department of Mechanical Engineering, University of Virginia, Charlottesville, Virginia 22903, USA}
\author{Abdullah Mamun}
\affiliation{Department of Electrical Engineering, University of South Carolina, Columbia, South Carolina 29208, USA}
\author{Kenny Huynh}
\affiliation{Department of Material Science and Engineering, University of California Los Angeles, Los Angeles, California 90095, USA}
\author{Marko Tadjer}
\affiliation{US Naval Research Laboratory, Washington DC, 20375, USA}
\author{Tatyana Feygelson}
\affiliation{US Naval Research Laboratory, Washington DC, 20375, USA}
\author{Karl Hobart}
\affiliation{US Naval Research Laboratory, Washington DC, 20375, USA}
\author{Travis Anderson}
\affiliation{US Naval Research Laboratory, Washington DC, 20375, USA}
\author{Bradford Pate}
\affiliation{US Naval Research Laboratory, Washington DC, 20375, USA}
\author{Alan Jacobs}
\affiliation{US Naval Research Laboratory, Washington DC, 20375, USA}
\author{James Spencer Lundh}
\affiliation{US Naval Research Laboratory, Washington DC, 20375, USA}
\author{Mark Goorsky}
\affiliation{Department of Material Science and Engineering, University of California Los Angeles, Los Angeles, California 90095, USA}
\author{Asif Khan}
\affiliation{Department of Electrical Engineering, University of South Carolina, Columbia, South Carolina 29208, USA}
\author{Patrick Hopkins}
\affiliation{Department of Mechanical Engineering, University of Virginia, Charlottesville, Virginia 22903, USA}
\author{Samuel Graham}
\affiliation{Department of Mechanical Engineering, University of Maryland, College Park, Maryland 20742, USA}
\email{samuelG@umd.edu}

\begin{abstract}
This study investigates thermal transport across nanocrystalline diamond/AlGaN interfaces, crucial for enhancing thermal management in AlGaN/AlGaN-based devices. Chemical vapor deposition growth of diamond directly on AlGaN resulted in a disordered interface with a high thermal boundary resistance (TBR) of 20.6 $m^2$-$K/GW$. We employed sputtered carbide interlayers (e.g., $B_4C$, $SiC$, $B_4C/SiC$) to reduce thermal boundary resistance in diamond/AlGaN interfaces. The carbide interlayers resulted in record-low thermal boundary resistance values of 3.4 and 3.7 $m^2$-$K/GW$ for Al$_{0.65}$Ga$_{0.35}$N samples with $B_4C$ and $SiC$ interlayers, respectively. STEM imaging of the interface reveals interlayer thicknesses between 1.7-2.5 nm, with an amorphous structure.  Additionally, Fast-Fourier Transform (FFT) characterization of sections of the STEM images displayed sharp crystalline fringes in the AlGaN layer, confirming it was properly protected from damage from hydrogen plasma during the diamond growth. In order to accurately measure the thermal boundary resistance we develop a hybrid technique, combining time-domain thermoreflectance and steady-state thermoreflectance fitting, offering superior sensitivity to buried thermal resistances. Our findings underscore the efficacy of interlayer engineering in enhancing thermal transport and demonstrate the importance of innovative measurement techniques in accurately characterizing complex thermal interfaces. This study provides a foundation for future research in improving thermal properties of semiconductor devices through interface engineering and advanced measurement methodologies.

\end{abstract}

\maketitle

\newpage
\section{ Introduction}

Extreme bandgap (EBG) semiconductors, like Al$x$Ga${1-x}$N ($x > 0.6$, with a bandgap energy $E_G > 4.88$ eV), have garnered attention for their potential applications in high voltage, high power, and high-temperature power electronics.\cite{Kaplar2016,Baca2020,Nanjo2013,Tsao2017} However, optimal device performance and power output in AlGaN-based devices, such as High Electron Mobility Transistors (HEMTs), hinges on efficient thermal management.\cite{Nigam2017,Bothra2017}  The inherent low thermal conductivity of AlGaN (approximately 8-10 W/m-K) and significant thermal resistance of adjacent interfaces within the HEMT structure results in enhancing hot spots generated within the channel on the drain-side of the gate.\cite{Gu2021,Gerrer2021,Felbinger2007} Inefficient substrate-side heat removal towards a heat-sunk device package can improve through the implementation of heat spreading layers, directly integrated atop the device, offering alternative pathways for heat dissipation in close proximity to the hot spot.\cite{Nigam2017,Bothra2017} The intrinsic high thermal conductivity of diamond makes it a prime candidate for top-side heat spreading layers. For example, Malakoutian et al. demonstrated for a GaN-on-Al$_2$O$_3$ device, that the inclusion of 650 nm polycrystalline diamond grown on-top of the device could reduce channel temperatures by $\leq$42\% at 6 W/mm.\cite{Malakoutian2021-3} Arivazhagan et al. similarly showed with thermal simulations that a GaN-on-Si HEMT device could match the operating temperatures of an equivalent GaN-on-SiC device if a diamond top-side heat spreader was implemented, emphasizing the potential benefits of optimizing diamond heat spreaders.\cite{Arivazhagan2021} However, achieving diamond integration without affecting the device's electrical performance requires careful consideration.

Integrating diamond top-side heat spreaders on HEMTs presents several challenges. In AlGaN HEMTs, the top surface can be etched or damaged by the hydrogen used in the microwave plasma chemical vapor deposition (MPCVD) growth of diamond films, especially when the growth temperature exceeds 600°C.\cite{Babchenko2017} Babchenko et al. demonstrated that a 60-minute hydrogen plasma treatment on an AlGaN/GaN heterostructure resulted in an increase in gate leakage current by approximately six orders of magnitude, significantly impacting the electrical properties of the device.\cite{Babchenko2017}

 In terms of thermal interfaces with GaN, Yates et al. showed an interface resistance exceeding 20 $m^2$-$K/GW$ for CVD diamond grown directly onto GaN, which resulted in pitting of the GaN surface.  However, the use of layers such as a 5 nm thick silicon nitride were shown to yield low thermal boundary resistance on the order of 10 $m^2$-$K/GW$. The silicon nitride interlayer between diamond and GaN acts as a protective barrier against hydrogen etching, safeguarding the semiconductor surface during diamond growth. Additionally, these layers promote vibrational bridging and adhesion at the diamond/AlGaN interface, thereby reducing the thermal boundary resistance (TBR) of the interface and enhancing heat dissipation through the top-side heat spreader. Previous research demonstrated a low TBR of 3.1 $m^2$-$K/GW$ for a diamond/GaN interface by reducing the thickness of the Si$_3$N$_4$  layer to 1 nm, which shows the importance of controlling the interfacial layer thickness during processing. However, the implementation of carbide interlayers between diamond and AlGaN has not been explored to date and represents a primary objective of this work. Carbide interlayers show potential for acting as phonon bridging layers between AlGaN and diamond, due to phonon density of states overlap with the carbides. This is discussed in more detail in the results section of this paper.

A challenge associated with thermal characterization of the effectiveness of interfacial layers between diamond and AlGaN is the limitation imposed by the thermal penetration depth and contour uncertainty. The buried nature of this interface, situated beneath hundreds of nanometers of diamond, poses difficulties for thermal characterization techniques such as time-domain thermoreflectance (TDTR), which can become insensitive to thermal resistances at such depths, exceeding the measurement technique's thermal penetration depth. Moreover, there is the issue of contour uncertainty, wherein an increase in the number of unknown parameters being fitted leads to higher uncertainty in their fitted values. To determine the thermal boundary resistance (TBR) of the diamond/AlGaN interface, we must also fit the TBR of the transducer/diamond interface and the thermal conductivity of the grown diamond. Consequently, measuring buried thermal resistances without encountering significant uncertainty in the obtained value presents a challenge. However, by employing multiple measurement techniques with varying thermal penetration depths, we aim to mitigate this contour uncertainty and effectively measure buried diamond/AlGaN interfaces.

In this study, we aim to implement nanocrystalline diamond heat spreading layers on-top of AlGaN grown on AlN substrates with the use of carbide-based interlayers to improve thermal transport. Multiple different carbide interlayers are investigated (i.e., SiC, B4C, SiC/B4C) for two different Aluminum composition AlGaN (65\% and 83\% Al). Following Diamond growth, each of the interfaces were inspected via scanning tunneling electron microscope to measure the resulting carbide interlayer thickness and inspect the quality of the interface. Atomic force microscopy was used to measure the grain-size of the grown Diamond to provide insight to the obtained thermal conductivities for each Al concentration. Additionally, we pioneer a fitting scheme that leverages multiple thermal measurement techniques (SSTR and TDTR), enabling the accurate probing of buried thermal interface resistances—a task challenging to achieve with a singular measurement setup alone.

\section{Experimental Methods}
\begin{figure}
    \includegraphics[width=\linewidth]{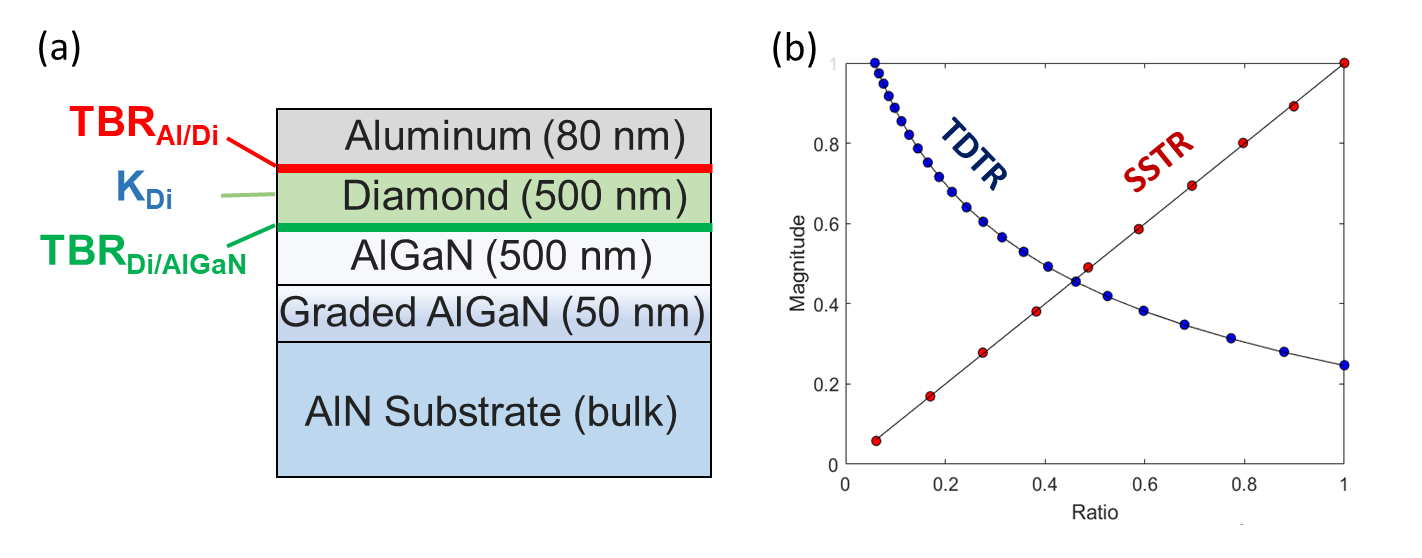}\par
    \caption{(a) Sample structures investigated in this study with nominal thickness values. (b) Example data set and fitting of hybrid SSTR-TDTR data.}
    \label{f-SampleHybrid} 
\end{figure}

The structure of the grown samples are shown in Figure \ref{f-SampleHybrid}(a). The epilayer structure for our study was deposited on physical vapor transport (PVT)-grown 400 um thick (0001) orientation and 0.25\degree miscut, commercially available 2-inch bulk AlN substrates (threading dislocation density, TDD of about 103 cm-2) from Hexatech.\cite{Hussain2023} All the growths were done on the Al-face of the substrate. 

Prior to epitaxy, the as-received chemical-mechanical polished (CMP) substrate was subjected to acid-based surface preparation, hydrogen annealing, and nitridation.\cite{Mamun2023} Subsequently, using MOCVD, epilayer structures as shown in Figure 1(a) were deposited over the AlN substrate. All the growths were carried out at 1100 – 1200 °C and 40 torr using trimethylaluminum (TMA), trimethylgallium (TMG), and ammonia (NH3) as the precursors. All the layers were undoped. AlN nucleation layers were grown at 1200 °C, whereas high-Al Al$_x$Ga$_{1-x}$N (x $>$ 0.65) were grown at 1160 °C. A growth rate of 150 nm/h was achieved at the AlGaN growth conditions. \cite{Washiyama2018}

The increasing lattice mismatch with increasing Ga-content in AlGaN alloy engenders large compressive stress of several GPa, resulting in large wafer bow for the growth of thick AlGaN epilayers on AlN substrates. Of the few reports on AlGaN growth on native AlN substrates, Dalmau et al. observed 8\% relaxation for a 400 nm thick Al0.65Ga0.35N layer.\cite{Dalmau2011} Grandusky et al. observed pseudomorphic growth of Al$_{0.6}$Ga$_{0.4}$N until a thickness of 0.5 µm. \cite{Grandusky2009} At larger thicknesses, they observed relaxation and an increase in TD density, which possibly formed via the interaction between misfit dislocations. Ren et al. reported the use of superlattice buffer layers for strain relief and found an inverse relationship between the degree of relaxation and surface roughness. \cite{Ren2007} In our case, we used linearly graded AlGaN as a potential strain relief layer while concomitantly maintaining the low TD density. 

For the characterization of the epilayers in this study, the AFM surface analysis was done using a Veeco instrument both for the starting AlN substrate (after surface preparation), and the as grown heterostructure. The AlN substrate shows uniform parallel steps with root mean square (RMS) roughness of about 0.1 nm. For thick AlGaN layers, the RMS roughness was 0.24 – 0.33 nm for different compositions. 

Boron carbide ($B_4C$) and silicon carbide ($SiC$) layers were sputtered in an AJA International sputtering system with $B_4C$ and $SiC$ targets indium-bonded to a copper base (Kurt J. Lesker). Both the $SiC$ and $B_4C$ layers had nominal thickness of 2 nm each. $SiC$ was sputtered at room temperature, 30 W/in2 RF power, 3 mTorr pressure, and 40 sccm Ar flow. $B_4C$ was sputtered at room temperature, 20 W/in2 RF power, 3 mTorr pressure, 40 sccm Ar flow. For the dual-layer deposition, the $SiC$ sputtering process was followed by $B_4C$ in situ for a nominal thickness of 2 nm for each layer. Ellipsometry measurement of this stack, referred to as $SiC$/$B_4C$ henceforth, deposited on a Si control sample, suggested a total thickness of 4.5 nm and a 2.04 refractive index. Prior to sputtering, the AlGaN samples were megasonic cleaned followed by etching in 10:1 dilute HCl for 30 seconds, 10:1 buffered oxide etchant for 60 seconds, and placed under vacuum less than 30 minutes after acid cleaning. Nanocrystalline diamond growth was performed in an Astex 1.5 kW microwave assisted plasma CVD reactor at 750 C, 800 W power, 15 Torr pressure, 900/3.0 sccm H2/CH4 gas ratio, resulting in a 0.52 um thick NCD layer. Additional characterization was performed via Nomarski microscopy, AFM imaging, and Raman spectroscopy.

We used two laser-based pump-probe measurement techniques for the measurement of thermal properties: Time Domain Thermoreflectance (TDTR) and Steady State Thermoreflectance (SSTR). In both techniques, a laser beam heats the sample (pump beam) while a second beam (probe) monitors the pump-induced temperature rise. In both cases, the data is fit to an analytical model for the transient heat equation for a multi-layered sample, allowing extraction of unknown thermal properties. In TDTR, pulsed beams are used, and the arrival of the pump and probe pulses is temporally offset, allowing reconstruction of a time-dependent thermal decay. In SSTR, the amplitude of the temperature rise is recorded under varying pump powers, where the slope can be considered roughly inversely proportional to the thermal resistivity within the measured region. Both techniques require the use of a thin metal transducer (80 nm aluminum in our case) to isolate the deposition of heat and probe thermoreflectance signal to the surface of the sample. Similarly, for SSTR, precise values for pump absorption or thermoreflectance coefficients are not required, as a reference sample with known properties and an identical surface (transducer material and surface finish) is used to account for these scaling terms. For our measurements, we use an additional reference sample to determine the substrate thermal properties, four-point probe electrical resistivity measurements to find the transducer thermal conductivity, and picosecond laser ultrasonics to determine the transducer thickness. Within the TDTR and SSTR thermal models, the material properties of each layer for the fabricated samples are shown in Table \ref{t-matprop}.
\begin{table}[t]
   \caption{Material properties for TDTR and SSTR analysis.}

    \setlength\tabcolsep{5pt} 
    \small
    \centering
    \begin{tabular}{cccccc}
        \toprule \toprule 
        
        Layer & Material & Thickness (nm) & \textit{k} (W m\textsuperscript{-1} K\textsuperscript{-1}) & \textit{c\textsubscript{p}} (J kg\textsuperscript{-1} K\textsuperscript{-1}) & $\rho $ (g cm\textsuperscript{-3}) \\ 
        \midrule \addlinespace[0.5em]
        
        \multirow{5}{*}{Film} & Al & 81.2\rpm2.1  & 121.0\rpm6.1 & 126.0\rpm3.0 & 19.3\rpm0.48 \\ 
        \addlinespace[0.5em]
        
        & Diamond & 500\rpm30.0  & Fitted & 520.0 \rpm26.1 & 3.51\rpm0.18 \\ 
        \addlinespace[0.5em]
        
        & AlGaN & 500\rpm26.5  & 9.2\rpm10.2 & 449.0\rpm11.1 & 7.19\rpm0.18 \\ 
        \addlinespace[0.5em]

         &Graded AlGaN & 50\rpm5.5  & 9.2\rpm10.2 & 449.0\rpm11.1 & 7.19\rpm0.18 \\ 
        \addlinespace[0.5em]
        \bottomrule
        \addlinespace[0.5em]
          
        Substrate &AlN    & 1.0 x 10\textsuperscript{6}   &321\rpm10.2 &  451.0\rpm11.2 & 3.28\rpm0.08\\ 

        \bottomrule \bottomrule
           \label{t-matprop}
    \end{tabular}
\end{table}

In our Al/Diamond/AlGaN/AlN samples, our primary focus was on determining the effective thermal boundary resistance of the Diamond/Carbide/AlGaN junction ($TBR_{Di/AlGaN}$), as labeled in Figure \ref{f-SampleHybrid}. However, due to the thickness of the Diamond (size effects) and its crystallinity (defect effects), we also needed to determine the thermal conductivity of Diamond ($K_{Di}$) and the thermal boundary resistance of the Aluminum/Diamond interface ($TBR_{Al/Di}$). The sensitivity of each measurement technique employed in this study to these unknown thermal parameters is compared in the supplemental information. It is worth noting that both techniques exhibit sensitivity to all three aforementioned parameters. 

\begin{figure}
    \includegraphics[width=1.0\linewidth]{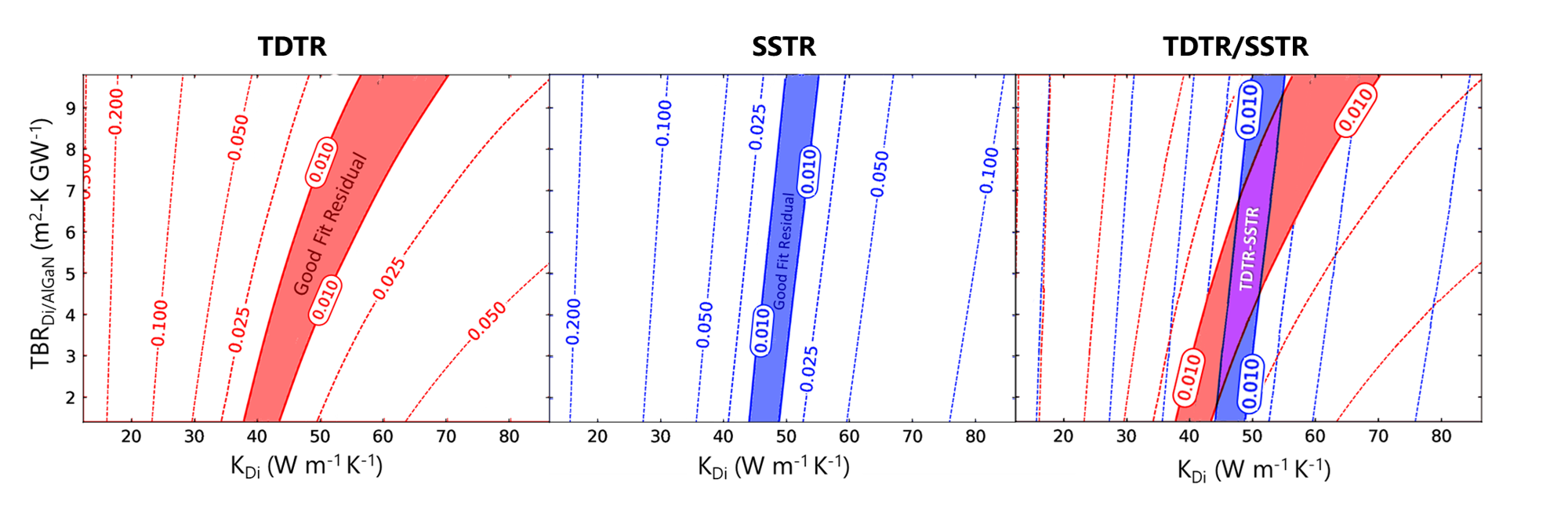}\par
    \caption{Two-dimensional contour plots for $TBR_{Di/AlGaN}$ and $K_{Di}$ shown for TDTR (red) and SSTR (blue), and our newly discovered hybrid technique of TDTR/SSTR. For each contour plot, the regions which provide a good fit to the experimental data (i.e., residual value $<$ 0.01) are shaded and labeled.}
    \label{f-2DContour} 
\end{figure}

Attempting to fit for all three unknown parameters using only TDTR data poses a significant challenge. A wide range of values for the thermal boundary resistance ($TBR_{Di/AlGaN}$) can lead to acceptable fits to the measured data. To identify and visualize these acceptable values, we explore various ranges for the unknown parameters of the system and calculate the residual between the curve produced by each parameter combination and the measured data. A threshold for the quality of the fit is selected, such as 1\% (0.010) in the case of exceptionally clean data, to constrain the acceptable values derived from our measurements. Conceptually, for the three unknown parameters, this process results in a 3D volume of acceptable fitted values. However, for simplicity, we will hold $TBR_{Di/AlGaN}$ constant and graph the range of acceptable values for $K_{Di}$ and $TBR_{Di/AlGaN}$, representing a 2D slice shown in Figure \ref{f-2DContour}.

Notably, the contours plotted for SSTR and TDTR, shown in Figure \ref{f-2DContour}, exhibit different shapes. Unfortunately, for the sample structures in this study, described in Figure \ref{f-SampleHybrid}(a), we find that the acceptable fit regions for both TDTR and SSTR (residual$<$0.010) span a large range of $TBR_{Di/AlGaN}$ and $K_{Di}$ values. This indicates that the contour uncertainty of our measurements would be large if we solely used either TDTR or SSTR. However, inspecting the 2D contour graph labelled TDTR/SSTR in Figure \ref{f-2DContour}, we notice a smaller area of intersection where the acceptable fit regions of SSTR and TDTR overlap. This indicates that, by simultaneously fitting both TDTR and SSTR measured data using a hybrid fitting scheme, the range of acceptable fitted values obtained is limited to those that yield acceptable fits for both SSTR and TDTR datasets (the purple intersected region). An example dataset and fit are illustrated in Figure \ref{f-SampleHybrid}(b). Employing this hybrid fitting approach significantly reduces the contour uncertainty in the fitted values of our unknown parameters. A similar principle is applied when others fit to the ratio and magnitude of TDTR measurements or during multi-frequency TDTR measurements.

The thermal model and numerical fitting algorithm are tasked with determining a combination of three unknowns that yields an acceptable fit for both TDTR and SSTR datasets. The fitting process utilizes existing numerical tools, such as Python's scipy minimize function, where the residual is calculated independently for each dataset and model. The maximum residual, representing the worst fit between SSTR and TDTR, is then passed to the minimization function to ensure that the worst-fitting dataset drives the overall fit. Additionally, we adopt the residual value approach, as outlined in Cahill et al. (i.e. 1\% residual), instead of using a mean squared error, to avoid potential issues associated with differently weighting datasets due to variations in data scaling or dataset lengths.

It is important to note that this fitting procedure differs slightly from that used for fitting TDTR data collected at multiple modulation frequencies. In the case of multi-frequency TDTR fitting, the datasets can be directly appended to each other, or a surface-fitting algorithm can be employed. While this approach is also feasible for TDTR/SSTR hybrid fitting, it is essential to consider the scaling of the SSTR data to ensure compatibility with the TDTR datasets.

\section{Results and Discussion}
\subsection{Thermal Boundary Resistance of Carbide Interlayers}
\begin{figure*}
    \includegraphics[width=1.0\linewidth]{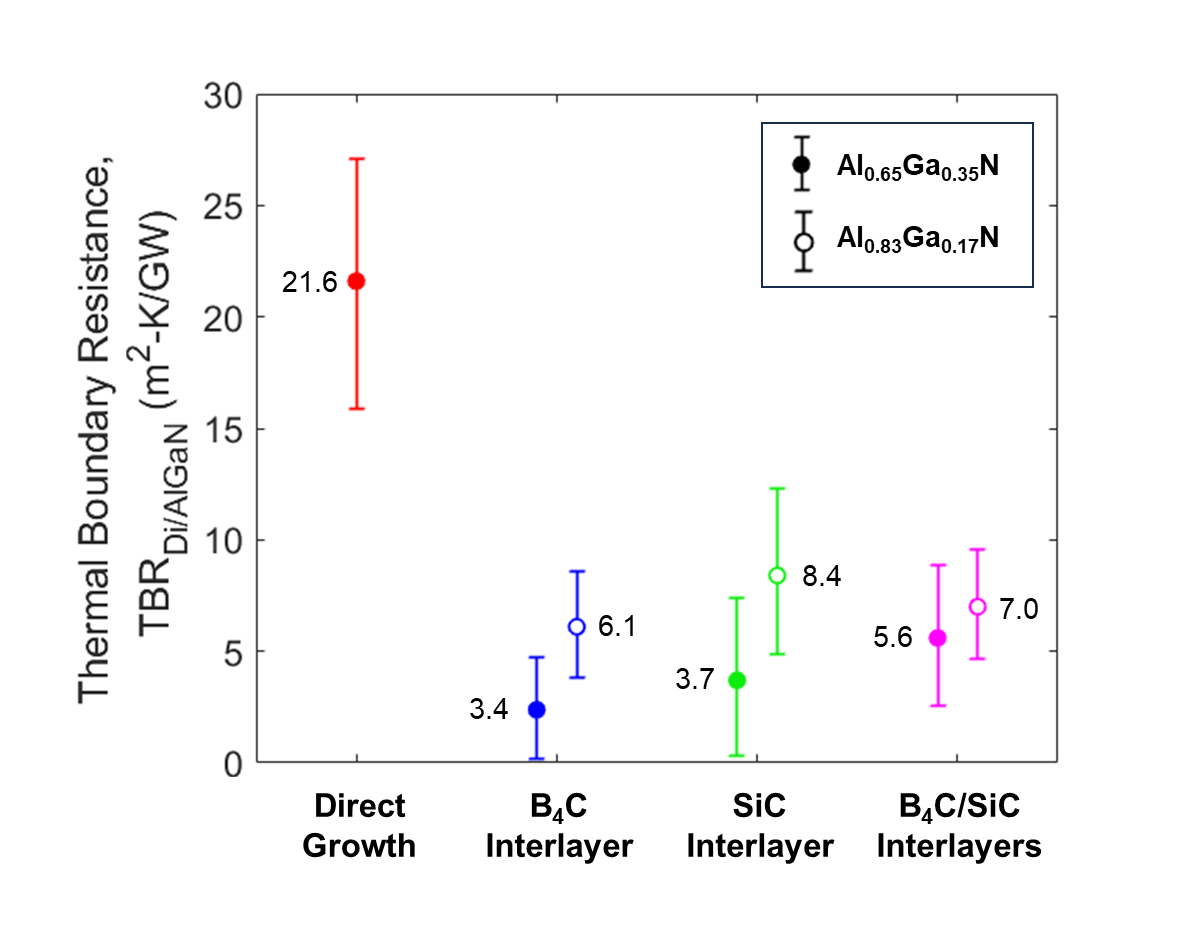}\par
    \caption{SSTR/TDTR measured results for $TBR_{Di/AlGaN}$ for each of the different samples. Direct growth is the case of no interlayer, with diamond grown directly on the AlGaN top layer. We notice all carbide interlayers significantly reduce $TBR_{Di/AlGaN}$ from the Direct Growth case, with $B_4C$ interlayer providing the lowest  $TBR_{Di/AlGaN}$ of 3.4 $m^2$-$K/GW$, a new record for Diamond/AlGaN interfaces.}
    \label{f-TBR} 
\end{figure*}

\begin{figure*}
    \includegraphics[width=\linewidth]{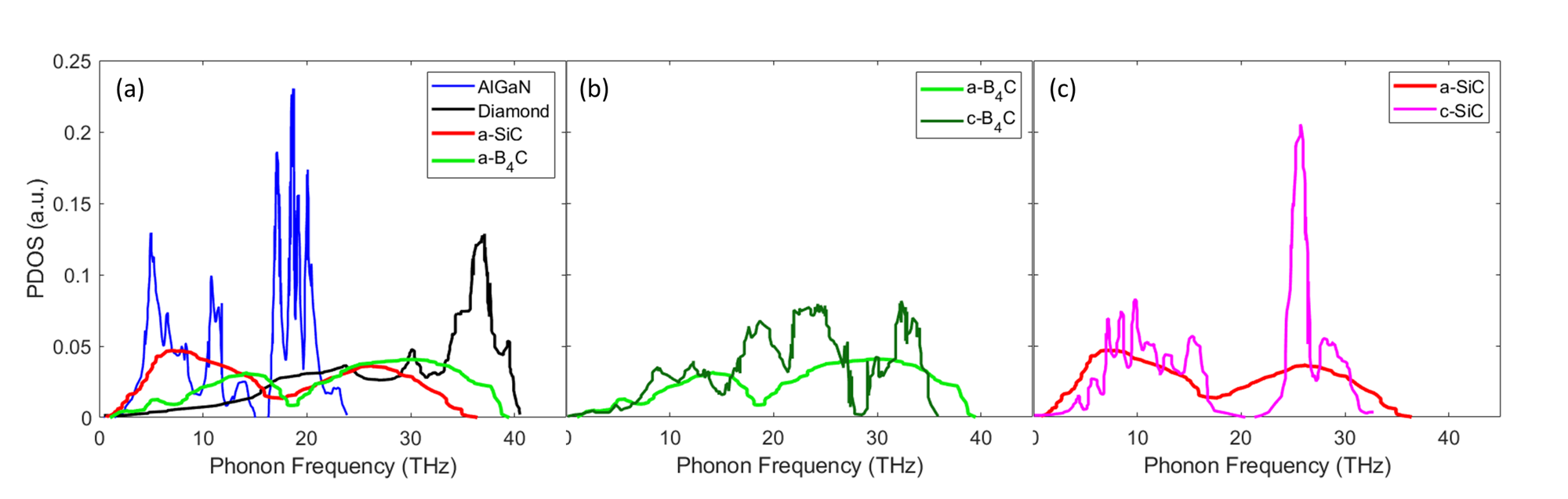}\par
    \caption{Graphs displaying the phonon density of states versus phonon frequency for the materials in our samples (a) Crystalline AlGaN, crystalline Diamond, amorphous $SiC$, and amorphous $B_4C$. $B_4C$ and $SiC$ both provide vibrational bridging via introducing intermediate phonon modes between AlGaN and Diamond. (b) A comparison between the crystalline and amorphous $B_4C$. (c) A comparison between crystalline and amorphous $SiC$.}
    \label{f-DOS} 
\end{figure*}

\begin{figure*}
    \includegraphics[width=1.0\linewidth]{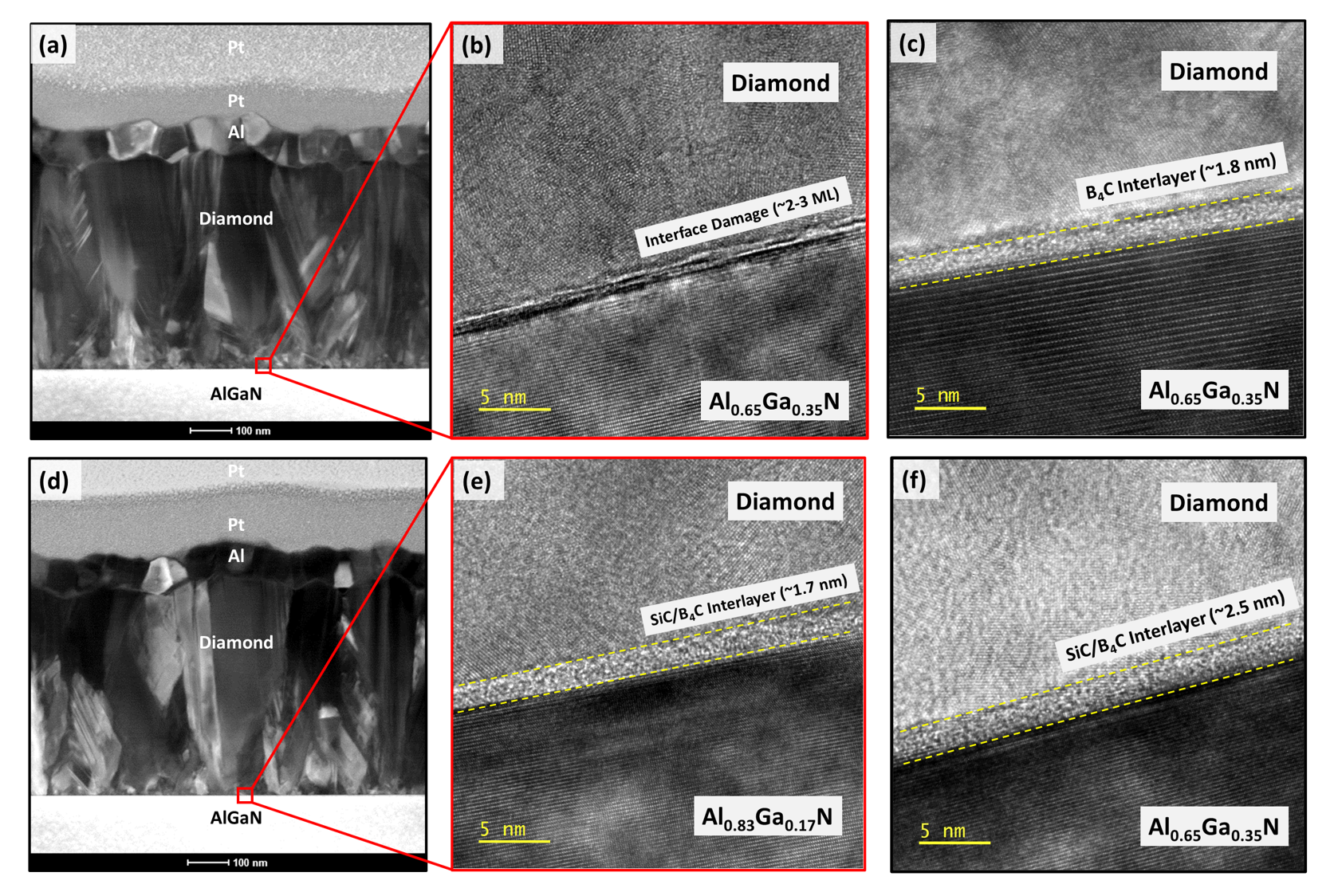}\par
    \caption{STEM images of the samples investigated in this study. (a) and (d) show the zoomed out STEM image for  65\% and 83\% Aluminum concentration AlGaN samples. (b) STEM image shows the Direct Growth sample (no interlayer) where there is visible structural disorder at the diamond/AlGaN interface. (c,e,f) STEM images of three interlayers, with the thickness of the interlayer shown. Notably, all interlayers are amorphous.}
    \label{f-STEM} 
\end{figure*}

The measured values of $TBR_{Di/AlGaN}$ from simultaneous fitting of SSTR/TDTR measurements are presented in Figure \ref{f-TBR}. The fitted values of $TBR_{Al/Di}$ for each sample were consistently found to be 28.2$\pm$1.5 $m^2$-$K/GW$, matching TBR values found in literature. Meanwhile, the fitted values $K_{Di}$  for both Aluminum concentrations of AlGaN (i.e., 65\% and 83\% Aluminum) were found to be 49.4 (error bars: 46.7 to 56.3) 74.9 (error bars: 66.4 to 81.4) $W/m-K$, respectively.

Direct MPCVD Diamond growth on AlGaN yielded a thermal boundary resistance ($TBR_{Di/AlGaN}$) of 21.6 $m^2$-$K/GW$. This substantial value can be attributed, in part, to the weak van-der-Waals bonding between Diamond and AlGaN and the absence of phonon frequency overlap between the two materials, as illustrated in Figure \ref{f-DOS}(a). Furthermore, as shown in Figure \ref{f-STEM}(b), STEM imaging revealed structural damage and disorder at the Diamond-AlGaN interface, likely caused by high-temperature H-plasma exposure damaging the AlGaN surface. Structural disorder at the interface creates scattering sites for phonons, which in turn increases the thermal resistance of the interface. 

\begin{figure*}
    \includegraphics[width=1.0\linewidth]{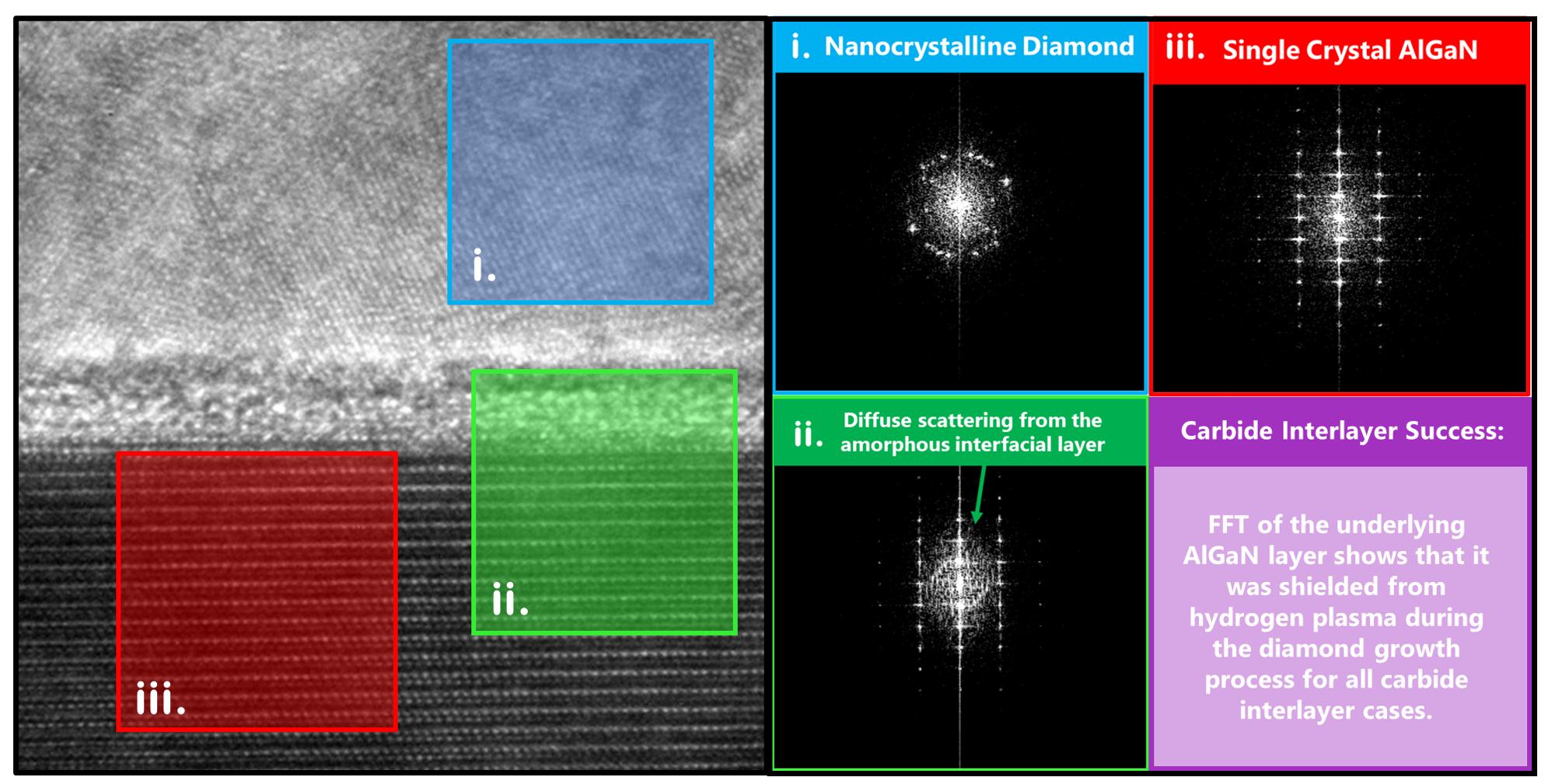}\par
    \caption{STEM image of the $B_4C$ sample investigated in this study, with the FFT characterization of separate section of the STEM image shown on the right. Notably, we find that all of our carbide interlayer samples protected the underlying crystalline AlGaN layer from hydrogen plasma used during diamond growth, due to its strong crystalline pattern in the FFT characterization.}
    \label{f-FFT} 
\end{figure*}

It is noteworthy that the inclusion of carbide interlayers (i.e., $B_4C$, $SiC$, and $B_4C/SiC$) effectively reduce $TBR_{Di/AlGaN}$ for all samples, shown in Figure \ref{f-TBR}. Remarkably, we measured a $TBR_{Di/AlGaN}$ of 3.4 $m^2$-$K/GW$ and 3.7 $m^2$-$K/GW$ for Al$_{0.65}$Ga$_{0.35}$N samples via the inclusion of $B_4C$ and $SiC$ interlayers, respectively, a new record for diamond/AlGaN interfaces. For the Al$_{0.83}$Ga$_{0.17}$N samples, we measured 6.1 $m^2$-$K/GW$ and 8.4 $m^2$-$K/GW$ for $B_4C$ and $SiC$ interlayers, respectively. This reduction is attributed to several factors. First, FFT characterization of the AlGaN layer in the STEM images, shown in Figure \ref{f-FFT}(iii), displayed distinct crystalline fringes, providing evidence that the carbide interlayers successfully protected the underlying AlGaN layer from damage from hydrogen plasma during the diamond growth. The lack of structural disorder in the AlGaN layer for the carbide interlayer samples would reduce the thermal resistance from the original base case (i.e., no interlayer). Second, the carbide interlayers provide vibrational bridging for phonons transporting between Diamond and AlGaN via the introduction of intermediate frequency phonon modes, as seen by their phonon density of states in Figure \ref{f-DOS}(a). Additionally, FFT characterization of the STEM images, shown in Figure \ref{f-FFT}(ii), revealed all carbide interlayers to be amorphous and have a thickness between 1.7-2.5 nm. Notably, amorphous materials are less confining on the phonon modes than their crystalline counterparts, enabling phonon mode conversions and broadening of the phonon density of states (improving PDOS overlap with Diamond and AlGaN), seen by the amorphous/crystalline comparison in Figure \ref{f-DOS})(b-c).\cite{Yang2018,Beltukov2018,Lv2016} Both of these factors facilitate greater phonon passage across the Diamond/Carbide/AlGaN junction, lowering $TBR_{Di/AlGaN}$. We speculate that amorphous $B_4C$ interlayers provided the lowest $TBR_{Di/AlGaN}$ values of all samples due to having the better phonon density of states overlap with diamond at high phonon frequencies than amorphous $SiC$, seen in Figure \ref{f-DOS}(a). 

\subsection{Advancements in Measuring Buried Thermal Boundary Resistances}
\begin{figure*}
    \includegraphics[width=1.0\linewidth]{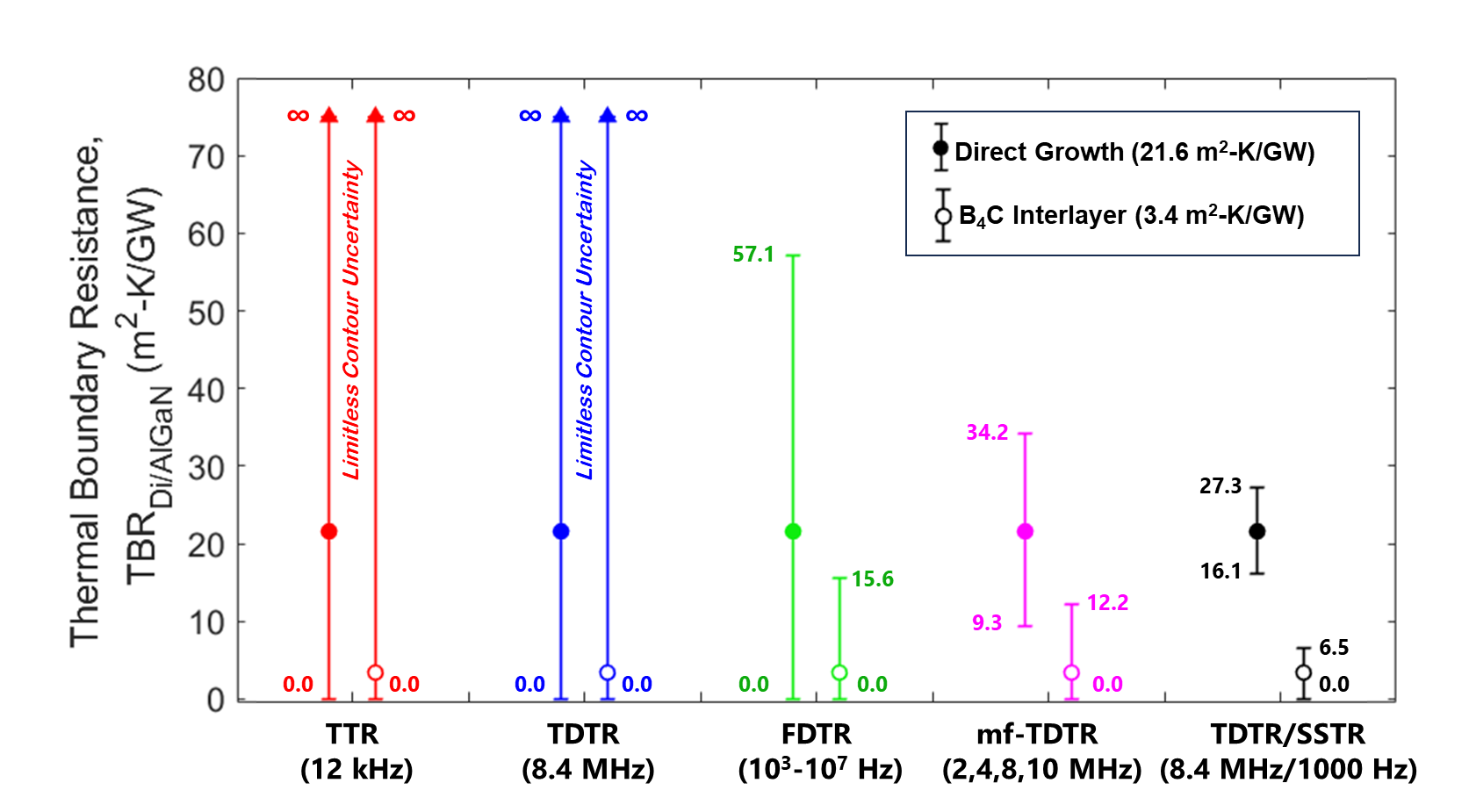}\par
    \caption{The measured value of $TBR_{Di/AlGaN}$ using various combinations of thermal measurement techniques, with the error bars quantifying the contour uncertainty associated with each fit.}
    \label{f-Contour} 
\end{figure*}

The thermal measurements discussed above are groundbreaking, not only shedding light on strategies for implementing diamond layers atop AlGaN devices (via carbide interlayers), but also by capturing buried thermal resistances that conventional thermoreflectance techniques struggle to detect. To underscore this significance, we present an analysis comparing the contour uncertainty associated with prominent thermoreflectance methods documented in the literature with our experimental measurements. Specifically, we examine the direct-growth Diamond/AlGaN sample (no interlayer, the largest $TBR_{Di/AlGaN}$ observed in our study) and the B4C interlayer sample (65$\%$ Al concentration in AlGaN, the smallest $TBR_{Di/AlGaN}$ observed in our study), with comparisons illustrated in Figure \ref{f-Contour}.

First, Transient thermoreflectance (TTR) and Time-domain thermoreflectance (TDTR) techniques both failed to measure $TBR_{Di/AlGaN}$ due to a phenomena that we here define as "Limitless Contour Uncertainty." Using TTR or TDTR systems to measure any of the samples in this paper will result in datasets with an unlimited possible combinations of fitted values for $TBR_{Al/Di}$, $K_{Di}$, and $TBR_{Di/AlGaN}$ that can produce a curve fitting the data well (i.e., residual error $\le$ 1$\%$). This presents a silent but serious danger to the thermal characterization community. Users not quantifying contour uncertainty for their system, may attempt to fit their measured data, just to extract values for $TBR_{Al/Di}$, $K_{Di}$, and $TBR_{Di/AlGaN}$ which are purely arbitrary and meaningless. For such a dataset, it is still possible to calculate the uncertainty of fitted values by propagating uncertainty in modeling parameters. However, in a Limitless Contour Uncertainty system, the values of fitted parameters and their propagated uncertainties are still meaningless.

The temporal resolution of TTR is insufficient for measuring our samples due to its single-pulse response nature, resembling an exponential decay. Consequently, attempting to fit three parameters using a single exponential time constant is not feasible, resulting in a myriad of fitted parameter combinations producing low residual error curve fits to the data. 

Time-domain Thermoreflectance (TDTR) (8.4 MHz modulation frequency) is limited in thermal penetration depth (approximated via the equation,  $d_{p,z} = \sqrt{\frac{K_z}{\pi f_m C}} = 1.01 \mu m$). While $TBR_{Di/AlGaN}$ is within the thermal penetration depth of TDTR, making it slightly sensitive to its value, it's not a dominant thermal resistance. TDTR is more sensitive to shallow large-value thermal resistances, like the thermal resistance of the diamond layer and $TBR_{Al/Di}$.

Frequency Domain Thermoreflectance (FDTR) encompasses a spectrum of thermal penetration depths, ranging from the beam radii (3.5 $\mu$m) at low frequencies to $d_{p,z} = \sqrt{\frac{K_z}{\pi f_m C}}$ at high frequencies. However, the influence of each thermal property on phase data at low frequencies is minimal, resulting in limited sensitivity to buried thermal resistances like $TBR_{Di/AlGaN}$. Thus, we observe large contour uncertainty for FDTR measurements of $TBR_{Di/AlGaN}$.

Multi-frequency-TDTR is robust at gaining sensitivity to buried thermal resistances via measuring the temporal response at different frequencies and different thermal penetration depths. However, a limitation of mf-TDTR is there exists a lower limit to modulation frequency at which data can be collected and still be clean. As such, mf-TDTR fitting resulted in the second lowest contour uncertainty, unable to probe buried thermal resistances as effectively as SSTR.

Lastly, the hybrid technique of SSTR-TDTR introduced in this study capitalizes on the strengths of both measurement techniques. The SSTR measurement utilizes non-normalized magnitude data to fit at very low modulation frequencies (1000 Hz), resulting in deep thermal penetration depths and heightened sensitivity to buried thermal resistances. Meanwhile, the TDTR measurement utilizes a temporal response at a high modulation frequency (8.4 MHz), resulting in lower thermal penetration depths and sensitivity to thermal resistances near the surface of the sample. As a result, we find SSTR-TDTR hybrid fitting produced the smallest contour uncertainty of the five techniques mentioned. This study elucidates the challenges associated with contour uncertainty during the measurement of buried interfaces and pioneers a hybrid fitting scheme that combines SSTR-TDTR to mitigate these challenges. This methodological advancement holds significant promise for future thermal measurements involving samples with numerous unknown parameters, where contour uncertainty poses a formidable obstacle.
 
\section{Conclusions}

This study provides a comprehensive analysis of thermal transport in nanocrystalline diamond/AlGaN interfaces, revealing significant insights into interface engineering and measurement methodologies. The measured values of thermal boundary resistance ($TBR_{Di/AlGaN}$) for direct MPCVD Diamond growth on AlGaN were found to be substantial at 21.6 $m^2$-$K/GW$. Scanning Tunneling Electron Microscope (STEM) images of the interface revealed atomics disorder in the AlGaN layer, a result of damage from hydrogen plasma during diamond growth. The large phonon mismatch between AlGaN and Diamond, and the atomic disorder in the AlGaN layer is responsible for the large $TBR_{Di/AlGaN}$ we measured for the direct-growth sample. However, the incorporation of sputtered amorphous carbide interlayers (i.e., $B_4C$, $SiC$, and $B_4C/SiC$), to assist phonon vibrational bridging, led to a remarkable reduction in $TBR_{Di/AlGaN}$, with the lowest values measured at 3.4 and 3.7 $m^2$-$K/GW$ for Al$_{0.65}$Ga$_{0.35}$N samples with $B_4C$ and $SiC$ interlayers, respectively. These values represent a new record for diamond/AlGaN interfaces, underscoring the efficacy of interlayer engineering in enhancing thermal transport across these interfaces. Additionally, Fast-Fourier Transform (FFT) characterization of the AlGaN layer displayed sharp crystalline fringes, providing evidence that the carbide interlayers successfully protected the underlying AlGaN layer from damage by hydrogen plasma during diamond growth. Meanwhile, FFT characterization of the STEM images of the interlayers provided proof that the sputtered carbide interlayers were amorphous, which could assist phonon mode conversion across the interface and improve phonon density of states overlap with diamond compared to their crystalline counterpart.

Furthermore, this study introduces a novel hybrid technique of simultaneous SSTR-TDTR fitting, pioneering a method that combines the strengths of both techniques to mitigate potentially large contour uncertainty associated with conventional thermoreflectance techniques. This hybrid approach offers superior sensitivity to buried thermal resistances, as demonstrated by the smallest contour uncertainty observed in our measurements.

These findings highlight the novelty and significance of this study in advancing the understanding of thermal transport in diamond/AlGaN interfaces. The results not only provide practical strategies for improving thermal management in AlGaN-based devices but also demonstrate the importance of innovative measurement techniques in accurately characterizing complex thermal interfaces. Overall, this study sets a foundation for future research in enhancing thermal properties of semiconductor devices through interface engineering and advanced measurement methodologies.

\section{Acknowledgements}
H.T.A., T.W.P., A.M., K.H., M.G., A.K., P.H., S.G. would like to acknowledge the financial support from U.S. ONR MURI Grant No. N00014-18-1-2429.

\bibliography{bibliography1}

\end{document}